\documentclass[aps,pre,floats,twocolumn,showpacs,superscriptaddress,reprint]{revtex4-1}
\usepackage{graphicx,epsfig}
\usepackage{graphics,dcolumn,bm,epic, eepic,fleqn,float}
\usepackage{amssymb,amsmath,amsfonts,multirow,rotate,color,xcolor}
\usepackage{soul}
\usepackage{url}

\newcommand{\avg}[1]{\langle #1 \rangle}
\usepackage{subfigure}

\def\Erdos{Erd\"os}
\begin{document}
\title{Social and place-focused communities in location-based online
  social networks}
\author{Chlo\"e Brown}
\affiliation{Computer Laboratory, University of Cambridge, Cambridge (UK)}

\author{Vincenzo Nicosia}
\affiliation{Computer Laboratory, University of Cambridge, Cambridge (UK)}
\affiliation{School of Mathematical Sciences, Queen Mary University of London, London (UK)}

\author{Salvatore Scellato}
\affiliation{Computer Laboratory, University of Cambridge, Cambridge (UK)}

\author{Anastasios Noulas}
\affiliation{Computer Laboratory, University of Cambridge, Cambridge (UK)}

\author{Cecilia Mascolo}
\affiliation{Computer Laboratory, University of Cambridge, Cambridge (UK)}
\date{\today}

\begin{abstract}Thanks to widely available, cheap Internet access and the
  ubiquity of smartphones, millions of people around the world now use
  online location-based social networking services. Understanding the
  structural properties of these systems and their dependence upon
  users' habits and mobility has many potential applications,
  including resource recommendation and link prediction. Here, we
  construct and characterise social and place-focused graphs by using
  longitudinal information about declared social relationships and
  about users' visits to physical places collected from a popular
  online location-based social service. We show that although the
  social and place-focused graphs are constructed from the same data
  set, they have quite different structural properties. We find that
  the social and location-focused graphs have different global and
  meso-scale structure, and in particular that social and
  place-focused communities have negligible overlap. Consequently,
  group inference based on community detection performed on the social
  graph alone fails to isolate place-focused groups, even though these
  do exist in the network. By studying the evolution of tie structure
  within communities, we show that the time period over which location
  data are aggregated has a substantial impact on the stability of
  place-focused communities, and that information about place-based
  groups may be more useful for user-centric applications than that
  obtained from the analysis of social communities alone.
  \end{abstract}

\maketitle

Networks can describe a large variety of complex systems, and network
science has proved to be a successful framework for the quantitative
study of their structure and
dynamics~\cite{Barabasi2002rev,Newman2003rev,Boccaletti2006}. In the
last decade, the tools and models provided by complex network theory
have enabled discovery of similarities between seemingly very
different systems including the Internet, the human proteome, and
collaboration networks. Complex networks analysis is now regularly
employed to characterise the topology and functioning of biological,
technological and social structures~\cite{Barrat2008,Newman2010}.

The analysis of social networks is one of the traditional application
fields of network science, and sociologists generally agree that many
social behaviours, from opinion formation to rule enforcement, from
individual success to cooperation, depend in a fundamental way on the
structure and evolution of the patterns of social relationships. In
other words, characterising and quantifying social structures is often
a prerequisite for understanding and interpreting social
dynamics~\cite{granovetter_1973,coleman_1988,Castellano2009}.
In the last twenty years, sociologists have relied on the study of
small social networks with tens or hundreds of nodes at most,
collected by means of targeted questionnaires and direct interviews.
Recently, the ubiquity of the Internet and the World Wide Web, and the
emergence of hundreds of online social services, have produced a huge
volume of data about online relationships between millions of people
around the world. These online social networks (OSNs) have allowed
quantitative verification of sociological theories on an unprecedented
scale. Analysis of a wide variety of online social systems has allowed
insights into the dynamics of human behaviours, including bond
formation, cooperation, imitation, and
synchronisation~\cite{Szell2010,Onnela2010,Lewis2012}. However, the
extent of the correspondence between people's online activities and
their offline lives is still the subject of
debate~\cite{Borge-Holthoefer2011,Crandall10:Inferring,Jones2013}.

One problem of interest in social network analysis is that of
identifying communities, cohesive groups of people who are more
tightly interconnected to each other than to the rest of the network.
Communities can be exploited in a wide range of practical
applications, including obtaining coarse-grained visual
representations of large networks, sorting personal online contacts
into manageable groups, finding partitions to speed up the performance
of services or providing personalised
recommendations~\cite{Liu12:Simplifying,Papadopoulos11:Community,Pujol10:Little}.
Many methods have been proposed in the last decade to find the best
partition of a graph into a set of meaningful
communities~\cite{Girvan2002,Newman2004a,Newman2006,Fortunato2009},
and the community structure of OSNs has recently been the subject of
much
research~\cite{Backstrom06:Group,Kumar10:Structure,Mislove07:Measurement}.
At the same time, OSNs are becoming increasingly location-aware,
meaning that user-produced content has an associated place. Examples
include Facebook's recent introduction of the ability to tag any post
with a location~\cite{Facebook12:Building}, geo-tagged tweets in
Twitter~\cite{Twitter09:Think}, and explicitly location-based social
networks. The most popular of these, Foursquare, has almost 35 million
users~\cite{Foursquare13:What}. Recent research has shown that online
social ties are more likely to form between spatially close users than
between those further
apart~\cite{Crandall10:Inferring,Backstrom10:Find,Cranshaw10:Bridging,Liben-Nowell05:Geographic,Scellato11:Socio-spatial},
but the exact role played by space in the formation and evolution of
communities is still unclear.

In this work we study Gowalla, an online location-based social
network, and analyse friendship and co-location networks obtained from
longitudinal data corresponding to more than 3 months of activity by
around 150,000 users.  We focus on the structural properties and
evolution of social and local communities, defined as
tightly-connected groups of nodes in the social and in the co-location
graphs respectively, and find that in general the overlap between
social and local communities is small, if not negligible. A local
community is rarely a proper subset of a social community, and usually
contains members belonging to different social groups. Furthermore,
the probability of two unconnected nodes becoming connected is much
higher if they belong to the same local community, and users in the
same local community who have not been in the same place are more
likely later to visit a common place than users in the same social
community.  Finally, while the structure of the social communities is
relatively stable over time, local communities are more dynamic and
volatile. The differences between social and local communities
highlighted in this work provide a first piece of evidence that the
standard approach to social group inference, based on the detection of
communities in the social graph, can fail to capture the microscopic
dynamics of local groups. Our results suggest that information derived
from social community analysis should be appropriately complemented
with knowledge about individual activity before being used in
user-centric applications such as providing friend suggestions or
place recommendations.

\section{The data set}
\label{sec:1}
We analyse data from Gowalla, an online location-based social network
founded in 2009 and discontinued at the end of 2011, when the company
was bought by Facebook. The service allowed users to declare
friendship ties to other users, thus forming a social network. The
main user activity in Gowalla was the \textit{check-in}: users
indicating their presence at specific, named venues using a mobile
phone application. When users \textit{checked in} to places in
this way, geo-located and time-stamped records were stored in the
system, and their friends in the social network were notified of their
location.

\subsection{Data collection}
Our data set consists of a series of daily crawls of Gowalla downloaded
between $4^{th}$ May and $19^{th}$ August 2010, obtained using the
public API provided by Gowalla to allow other applications and
services to access their content. Each user is identified by an
anonymised numeric ID and has an associated profile including social
connections and check-ins.  We downloaded these profiles from the
service daily over the crawling period, meaning that for each day we
have complete information about the social graph (all the friendship
ties between users at that time), and about all the user check-ins.
Each check-in consists of the venue name, category, location (latitude
and longitude), the ID of the user who made the check-in, and a
timestamp. We also have a record of all the check-ins that had taken
place before the measurement period began, but we do not have the
state of the social network corresponding to this period.

In Gowalla each place is represented as a named venue, such as
`Starbucks', `Kings Cross Railway Station' or `Computer Laboratory',
with latitude and longitude values so that the correct `Starbucks' for
the user's location can be identified. The user therefore checks in to
a specific place, rather than being located using coordinates alone.
We can therefore identify when users actually visit \textit{the same
  places} rather than just being in geographic proximity, e.g., in two
shops next door to one another. Having crawled Gowalla daily, we are
able to examine closely which social ties were formed and deleted
during the data collection period, and gain insight into the dynamics
of the network structure at the level of individual links. The crawl
was performed when the network was already fairly large and steadily
growing, not during the explosive growth period typically observed
shortly after the creation of such online social services, when their
popularity increases exponentially~\cite{Kumar10:Structure}.

\begin{figure*}[ht]
  \begin{center}
    \includegraphics[width=6in]{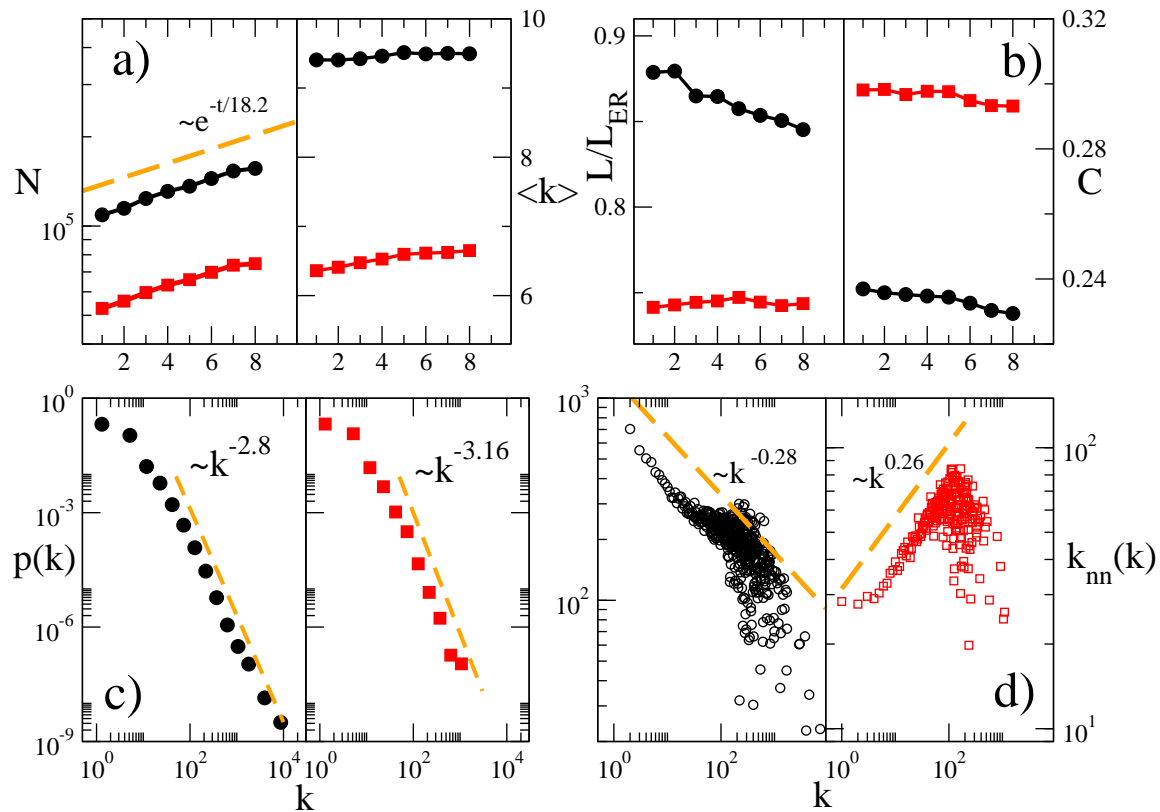}    
  \end{center}
  \caption{(colour online) Structural properties of the social (black
    circles) and placefriends graphs (red squares) over time. The
    number of nodes $N$ in both graphs increases exponentially with
    time (panel a, left), and the mean degree $\avg{k}$ increases only
    slightly (panel a, right). Panel b): the value of the mean
    shortest path length ($L$) divided by the expected value in a
    corresponding \Erdos-Renyi graph ($L_{ER}$) is smaller than those
    for the social and the placefriends graphs (left). Both networks
    have a relatively high clustering coefficient (right). Panel c):
    the tail of the degree distribution of the social graph (left,
    $\gamma\simeq 2.8$ in the social graph) is typically more
    heterogeneous than that of the placefriends graph (right,
    $\gamma\simeq 3.16$ in the placefriends graph). Panel d): due to
    the presence of many super-hubs, the average degree of first
    neighbours of a node having degree $k$ in the social graph is an
    increasing function of $k$ (left, disassortative degree
    distribution); conversely, the placefriends graph has assortative
    correlations (right). The results in panel c) and panel d)
    correspond to the whole observation interval.}
  \label{fig:fig1} 
\end{figure*}

\subsection{Data processing}

Since we have two kinds of information about users, i.e. the places
where they have checked in, and their connections in the social
network, given a time interval we can construct two different graphs.
The first graph $G = (V, E)$ represents the social network: each user
present in the system during the considered time interval is
represented by one of the $N=|V|$ nodes of the set $V$, and $E$ is a
set of $K$ edges (or ties) between nodes. The edge $(u_1, u_2)$ exists
in $E$ between users $u_1,u_2 \in V_i$ if $u_1$ and $u_2$ are friends
in the OSN in that time interval. We represent a graph by the
adjacency matrix $A=\{A_{ij}\}$, in which the entry $A_{ij}=1$ if
there exists a link connecting node $u_i$ and node $u_j$, and
$A_{ij}=0$ otherwise. The number of neighbours of a node $u_i$ is
called the degree of $u_i$, and is denoted by $k_i=\sum_{j}a_{ij}$. In
the following we refer to the average degree of a graph
$\avg{k}=2K/N$. In Gowalla, ties are bidirectional and
indistinguishable, so the social graphs we construct are undirected
and unweighted, and the associated adjacency matrices are symmetric.

Using the information about places where users check in, we can define
the notion of \textit{placefriends}: users (not necessarily having a
social tie) who have checked in to one or more of the same
places. Since our aim is to investigate the relationship between
online and offline social groups, we are particularly interested in
users who have checked in to one or more of the same places as their
online friends, that is, users who are both friends and placefriends.
Given a time interval and the corresponding social graph $G=(V,E)$, we
define the associated \textit{placefriends-social} graph $G^P = (V^P,
E^P)$ as the subgraph of $G$ such that $V^P$ contains all the nodes in
$V$ having at least one friend who is also a placefriend, and $E^P$ is
the subset of edges $(u_1, u_2)\in E$ such that $u_1$ and $u_2$ are
both friends and placefriends.  We call $N^{P}$ and $K^{P}$,
respectively, the number of nodes and the number of edges in the
placefriends graph. For convenience, we henceforth refer to the
placefriends-social graph simply as the \textit{placefriends graph},
and to the edges of the placefriends graph as \textit{placefriends edges
  (or ties)}.
  
\section{Structure of social and placefriends graphs}

We analyse the structure of the social and placefriends graphs in
Gowalla, focusing on the temporal evolution of communities. In order
to study the temporal evolution of these graphs, we divided the
original data set into 8 snapshots, each covering a period of 2 weeks
except the last one, which is 9 days long.
Table~\ref{tab:dataset_properties} reports the number of new check-ins
and new unique places per day at each snapshot, and the total number
of check-ins and unique places in the data set at the end of each
snapshot. The total number of check-ins in the first snapshot refers
to all check-ins recorded in the system since the inception of
Gowalla.  At the time of crawling the system was steadily growing,
with around 5,000 new places visited every day.

\begin{table}[thb]
\centering
\begin{tabular}{|c||c|c|c|c|}
\hline
Snapshot & NCh & NP & TCh & TP\\
\hline
\hline
1  & - & - & 4,946,778 & 1,023,991 \\
\hline
2  & 49,562 &  4,812 & 5,640,646 & 1,091,366 \\
\hline
3  & 59,055 &  4,997 & 6,467,416 & 1,161,324 \\
\hline
4  & 59,846 &  4,785 & 7,305,267 & 1,228,323 \\
\hline
5  & 52,085 &  4,876 & 8,034,466 & 1,296,594 \\
\hline
6  & 53,878 &  5,061 & 8,788,764 & 1,367,448 \\
\hline
7  & 57,941 &  4,921 & 9,599,945 & 1,436,352 \\
\hline
8  & 32,106 &  2,960 & 9,888,905 & 1,462,993 \\
\hline
\end{tabular}
\caption{The mean number of new check-ins per day (NCh), mean
  number of new unique places per day (NP), total number of
  check-ins (TCh), and total number of places (TP) in the data set at
  the end of each snapshot. The number of check-ins and the number of
  places grew steadily during the crawl.}
\label{tab:dataset_properties}
\end{table}

\subsection{Basic network properties}

Fig.~\ref{fig:fig1} reports the basic structural properties of the
social and placefriends graphs corresponding to each of the eight
snapshots. The number of nodes in the largest connected components of
both the social (black circles) and the placefriends graphs (red
squares) increases exponentially over time (Fig.~\ref{fig:fig1}a),
confirming that at the time of data acquisition the system was still
growing. In particular, the number of users in the largest connected
component of the social graph increased from around 100,000 to around
150,000, while the size of the largest connected component in the
placefriends graph grew from around 52,000 to around 75,000. The
average node degree in the social graph (right panel of
Fig.~\ref{fig:fig1}a) remains almost constant over the 8 snapshots,
indicating that the social network was already well-established and
stable at the time of the crawl. Conversely, the average degree of the
placefriends graph increases from around 6.2 in the first snapshot to
around 6.8 in the last, showing that the placefriends graph becomes
denser with time.

Both the social and the placefriends graphs are small-world networks,
as confirmed by Fig.~\ref{fig:fig1}b, which reports the values of the
relative characteristic path length (the average distance between any
pair of nodes divided by the expected value of this quantity in an
\Erdos-Renyi graph with the same number of nodes and links) and the
mean node clustering coefficient (the mean percentage of closed triads
incident to a node). The relative characteristic path length in the
placefriends graph is smaller than that of the social graph, and the
average clustering coefficient of the placefriends graphs is
consistently higher than that of the social graph, indicating that on
average a node in the placefriends graph is surrounded by neighbours
who in turn have a high probability of having been to a common
place. This effect can be explained by observing that users of online
social networking services may add as friends people they meet only
occasionally, if ever, since creating and maintaining this kind of
online friendship does not involve any real cost or
effort. Conversely, in order to be considered placefriends, two users
have to have been to the same place, meaning that their activities are
focused around a certain geographical area. This makes it more
probable that pairs of their friends with whom they share a common
place could themselves have visited a common place in the same area.

\subsection{Degree distributions and degree correlations}

In Fig.~\ref{fig:fig1}c we show the degree distribution, i.e. the
probability $P(k)$ of finding a node having degree equal to $k$, of
the social and placefriends graphs corresponding to the whole data
set. The two degree distributions exhibit power-law tails,
i.e. $P(k)\sim k^{\gamma}$ for large $k$, indicating that the form of
the distribution is scale-invariant and that there is a non-negligible
probability of nodes having a large number of neighbours. However, the
tails of the two distributions are characterised by two different
values of $\gamma$~\footnote{The exponents of the tails of the degree
  distributions have been computed using the maximum likelihood
  estimator (MLE) for discrete power-law distributions, as indicated
  in Ref.~\cite{Clauset2007a}. The values of $k_{min}$ corresponding
  to the maximum likelihood fit are $k_{min}=79$ and $k_{min}=67$,
  respectively for social and placefriend graphs.}. Since the exponent
of a power-law degree distribution is an indirect measure of the
heterogeneity of the degrees, with larger exponents corresponding to
more homogeneous distributions, we can conclude that the node degrees
of the placefriends graph usually are more homogeneously
distributed. The maximum degree of the placefriends graph is much
smaller than that of the social graph ($k_{max}\simeq 1000$ in the
placefriends graph while $k_{max}\simeq 10,000$ in the social
graph). These observations are consistent with the fact that
place-friendship is much more costly and demanding than purely online
friendship.

For complex networks, the degree distribution alone is often not
enough to fully characterise the microscopic structure.  Many networks
exhibit degree-degree correlation, meaning that the existence of a
link between two nodes having respective degrees $k$ and $k'$ is a
function of both $k$ and $k'$~\cite{Newman2002,Newman2003}. Networks
can be either \textit{assortative} (nodes of a certain degree are
preferentially linked to nodes with similar degrees) or
\textit{disassortative} (highly connected nodes are preferentially
linked with other nodes having small degree, and vice versa). The
assortativity of a network can be quantified by looking at the average
degree $k_{nn}(k)$ of the first neighbours of nodes having degree $k$,
as a function of $k$. For assortative networks, $k_{nn}(k)$ is an
increasing function of $k$, while for disassortative networks
$k_{nn}(k)$ decreases with $k$. Quite often, $k_{nn}(k)$ is a
power--law, i.e. $k_{nn}(k)\sim k^{\nu}$; in these cases, the exponent
$\nu$ can be used to quantify the assortativity of the network, with
more positive values of $\nu$ indicating more assortative networks and
more negative values of $\nu$ corresponding to disassortative graphs.

Fig.~\ref{fig:fig1}d reports the value of $k_{nn}(k)$ for the social
and the placefriends graphs. Notice that while the social graph is
markedly disassortative ($\nu\simeq -0.28$), the placefriends graph is
assortative ($\nu\simeq 0.26$). This means that hubs in the social
graph preferentially link to poorly-connected nodes, while nodes in
the placefriends graph tend to be connected with other nodes having
similar degree. We hypothesize that the disassortativity of the social
graph may be due to the nature of Gowalla as an online social service:
its most active users, who would probably add the most friends, tended
to be `early adopters' and people who were particularly interested in
new online services and technology. These people would have a lot of
connections and might convince their friends to sign up to the
service, but these friends could be less interested and maybe only add
one or two friends before stopping using Gowalla. Such patterns of
behaviour could give rise to the kind of disassortativity we see in
the social graph.

The results reported in Fig.~\ref{fig:fig1} confirm that the structure
of the social graph constructed from friendship declared by Gowalla
users is fundamentally different from the structure of the
corresponding placefriends graphs obtained from check-in
information. This means that social ties alone are probably not a good
proxy of users' activity, and that information about friendships needs
to be appropriately complemented with other knowledge before being
used to draw conclusions about users' dynamics.

\section{Social and local communities}
We have seen that despite being constructed from the same data set,
the social and placefriends graphs are quite different with respect to
heterogeneity and assortativity. We now focus on the community
structures of the two graphs, in order to understand whether these
discrepancies also reflect a different meso-scale organisation.  In
general, a community is a subset of nodes of a graph that are tightly
connected to each other. Depending on the precise definition of
community employed~\cite{Fortunato2009}, one can require that in order
to form a proper community a subset of nodes should either be more
tightly connected than expected in a null model~\cite{Newman2006} or
should instead have more internal links, i.e.~edges between nodes
belonging to the community, than external ones, i.e.~those connecting
a node inside the community with a node outside the
community~\cite{Girvan2002}. We use the former definition, and we
consider partitions obtained using the Louvain
method~\cite{Blondel08:Fast}, a greedy agglomerative community
detection algorithm based on modularity
optimisation~\cite{Newman2006,Arenas2007}.

It has been observed that some OSNs contain groups of users who are
online friends and also happen to visit the same places in the
physical world~\cite{Brown12:Importance}; in practice, these are
groups of placefriends who also form a community in the social
graph. Since we are interested in understanding the relationship
between online and place-focused communities, and in particular in
quantifying the extent to which the structure of the social graph
mirrors the activity of users visiting the same places, we will
compare the results of community detection performed on the Gowalla
social graphs and on the corresponding placefriends graphs. In the
following we call the communities of the social graph
\textit{social-only communities}, or simply \textit{social
  communities}, and refer to the communities of the placefriends
graphs as \textit{local communities}.

When tracking the evolution of communities over time, we need to be
confident that changes in the communities in two different temporal
snapshots are due to the changing structure of the network, not to
peculiarities of the community detection algorithm. Many greedy
community detection methods, including the Louvain method, are
non-deterministic and can give different output depending on the order
of the input~\cite{Fortunato2009}. To address this problem, we adopted
the algorithm proposed by Kwak et al.~\cite{Kwak09:Mining}, which
works as follows. A chosen community detection algorithm able to
handle weighted graphs (e.g., the Louvain method) is run $M$ times on
the same network, with the input being given in a different,
randomised order each time, thus obtaining $M$ community partitions of
the graph. In principle, if the graph has a strong community structure
then these $M$ partitions should differ only for the community
placement of a relatively small subset of nodes. Then, the network is
re-weighted according to the frequency with which pairs of nodes have
been placed in the communities of each of the $M$
partitions. Specifically, the weight $w_{ij}$ of an edge connecting
nodes $i$ and $j$ is increased or decreased proportionally to the
number of times that $i$ and $j$ have been put in the same community
in each of the $M$ runs. The re-weighting procedure has the effect of
reinforcing more robust groupings over those appearing by chance or
due to a particular input ordering. The process is iterated on the
re-weighted network, until a consistent placement of nodes into
communities is obtained and all the partitions obtained in the $M$
runs of an iteration are identical. Although Kwak et al.~noted in
their paper that the convergence of their method cannot be guaranteed
for certain graphs, we did not encounter this problem for our
networks, and we were able to find a stable partition of each graph.

\subsection{Size distribution}

\begin{figure}[tbh]
\centering
\includegraphics*[width=\columnwidth]{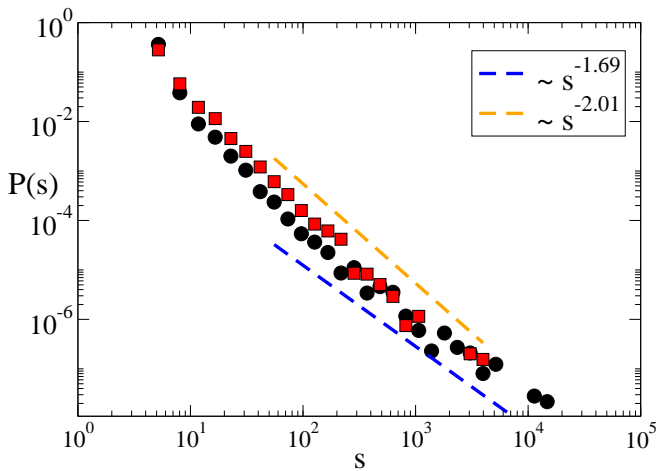}
\caption{(colour online) Distribution of community sizes of the social
  (black circles) and placefriends graphs (red squares) in the final
  snapshot of the data set. The tails of both distributions are
  power-laws, with exponents $\gamma=-1.69$ for social communities
  (solid blue line) and $\gamma=2.01$ for placefriends graphs (dashed
  orange line). This indicates that the size of large social
  communities is more heterogeneous.  Notice that placefriends
  communities are consistently smaller than social communities.}
\label{fig:sizes}
\end{figure}

It is common for social networks to exhibit communities at different
scales, and quite often the distribution of community sizes is a
power-law. Figure~\ref{fig:sizes} shows the size distributions of
communities in the final snapshot of the social and the placefriends
graphs in Gowalla; the distributions do not change significantly
between any of the snapshots. Notice that the distributions have
power-law tails, so that more than 95\% of both social and local
communities have fewer than $s=30$ members. However, the exponent of
the tail ($s>30$) of the distribution of social communities
($\gamma=1.69$) is smaller than that of the distribution of
placefriends communities ($\gamma=2.01$), indicating that the size of
large social communities is more heterogeneous~\footnote{The exponents
  of the tails of the distributions of community sizes have been
  computed using the maximum likelihood estimator (MLE) for discrete
  power-law distributions, as indicated in
  Ref.~\cite{Clauset2007a}. The fit was perfomed by setting
  $s_{min}=30$.}. We are particularly interested in small communities
when considering local communities: in a study of communities formed
by people communicating using mobile phones, Onnela et
al.~\cite{Onnela11:Geographic} found that communities of up to 30
people tended to be geographically tight, with the span of user
locations becoming larger more quickly once this size is
exceeded. Indeed, Figure~\ref{fig:sizes} confirms that the size of
placefriends communities is consistently smaller than that of social
communities. This might reflect the ease of establishing an online
tie, while the constraint that users within local communities have
been to the same places means that these communities are necessarily
smaller.

\subsection{Shared places}
\begin{figure}[tbh]
  \centering
  \includegraphics*[width=\columnwidth]{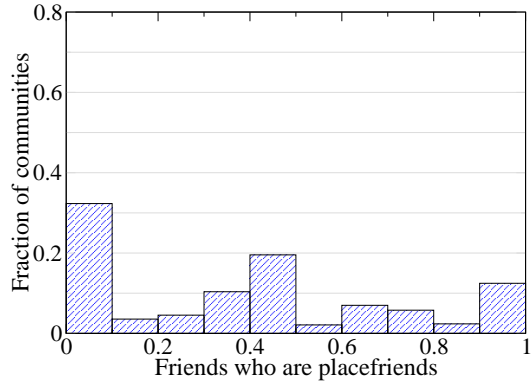}\\
  \includegraphics*[width=\columnwidth]{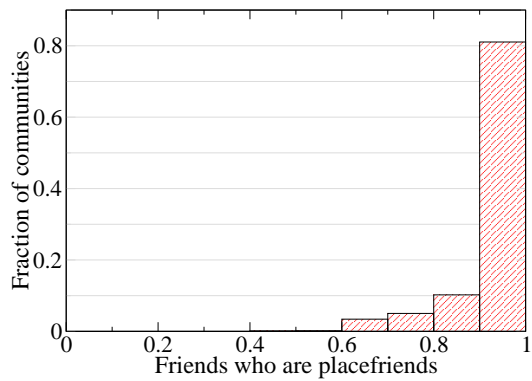}
  \caption{(colour online) Proportions of communities having particular
    fractions of intra-community social ties where connected users are
    also placefriends (have been to one of the same places), in the
    final snapshot of the social (top panel) and placefriends graph
    (bottom panel). Notice that more than $30\%$ of social communities
    have under $10\%$ of their connected nodes being
    placefriends, while more than $80\%$ of local communities contain
    have more than $90\%$ of their connected nodes being placefriends.}
  \label{fig:pffracs}
\end{figure}

Figure~\ref{fig:pffracs} shows how many intra-community social ties in
social and local communities are between placefriends (recall that
placefriends are two users who have been to one of the same places,
not necessarily having a tie in the social graph). Almost all of the
social ties between members of local communities connect users who are
also placefriends. In contrast, there are many social communities that
have no such ties.  For instance, we find that in more than 30\% of
the communities in the social graph, under 10\% of the internal edges
lie between nodes that are also placefriends, and in more than half
the social communities, under 50\% of the social ties are between
placefriends. In the local communities, however, we can see that in
over 80\% of the communities, more than 90\% of the internal social
ties are between placefriends.  We thus see that if we perform
community detection on the social graph alone we are not able to
identify many of the local communities.

Since Gowalla is an explicitly location-focused social network with
an emphasis on location sharing, one might expect users to be
friends with those who go to the same places. Indeed, this has been
shown to be the case by previous research into online location-based
social networks: Scellato et al.~\cite{Scellato11:Exploiting} found
that during the steady growth period of the service, 30\% of links
are added between placefriends. However, social communities mainly
contain ties between users who do not visit the same places.
This is a first piece of evidence that performing community
detection on the social graph may not capture local communities, even
though spatially close users are more likely to form ties than distant
users~\cite{Stewart41:Inverse}.

\section{Temporal evolution of communities}
We now study the formation and deletion of edges in the social network in
each type of community, by examining the social and local communities
present in the first snapshot, and identifying the social ties that
have been created and deleted by the time of the final snapshot. We
consider ties according to whether they are within or between social
and local communities. Note that the graph we use here is the subset
of the social graph $G$ composed of nodes that are in communities at
the first snapshot. Isolated pairs of nodes and nodes with no
ties are not considered to be in communities.

\subsection{Null model}
We compare the actual number of links created or deleted within or
between communities since the first snapshot, to the corresponding
expected number of links created or deleted in a null model, where new
links (resp.~old links) are added (resp.~removed) at random.  The only
constraint is that we avoid self-loops and multiple edges between the
same pair of nodes. If $x$ is the observable of interest (i.e.~number
of links meeting the given criteria), we denote by $\tilde{x}$ the
expected value of $x$ in the null model, and we compute:
\begin{equation}
  r = \frac{x}{\tilde{x}}
\end{equation}
For instance, if we want to assess the significance of edge creation
between nodes belonging to the same community, the observable $x$ is
the total number of edges added between nodes in the same community
since the first snapshot, while $\tilde{x}$ is the expected number of
edges created between nodes in the same community when the same number
of new edges are placed uniformly at random. That is, the total number
of edges added since the first snapshot, multiplied by the fraction of
those edges that could be formed between nodes belonging to the same
community. This quantity can be thought of as the number of `missing'
edges within communities, i.e.  the number of pairs in the same
communities that are not connected in the first snapshot. Thus, the
expression for $\tilde{x}$ reads:
\begin{equation*}
  \tilde{x} = \frac{K^{in}_{\circ}}{K_{\circ}} K_{+}
\end{equation*}
where $K^{in}_{\circ}$ is the number of edges missing between members
of the same community, $K_{\circ}$ is the total number of missing
edges (between members of the same community and members of different
communities), and $K_{+}$ is the number of edges added between the
first and the final snapshots.
Instead, if we consider intra-community edge deletion, $\tilde{x}$ is
equal to the total number of deleted edges in the final snapshot
multiplied by the fraction of edges that lie within a community in
the first snapshot. As a formula:
\begin{equation*}
  \tilde{x} = \frac{K^{in}_{\bullet}}{K_{\bullet}} K_{-}
\end{equation*}
where $K^{in}_{\bullet}$ is the number of edges between members of the
same community in the first snapshot, $K_{\bullet}$ is the total
number of edges in the first snapshot and $K_{-}$ is the number of
edges that have actually been deleted between the first and the final
snapshot.

The same model applies when considering the case of edges created or
deleted between communities, but with the obvious replacements of the
edge counts for within communities with those for between
communities. In summary, $r$ is the ratio of the number of
edges we actually see being created or deleted between members of the
same or different communities, to that we would expect if edges were
to be added or removed at random. We can use this to assess the
significance of communities for the creation and deletion of edges.

In Figure~\ref{fig:pretty_circles} we give a visual representation
of the ratios $r$ corresponding to the formation and deletion of edges
in the social and in the placefriends graph. In each panel of the figure we represent the results for social and local community communities (yellow and blue boxes respectively).

\subsection{Edge formation and deletion}
\begin{figure*}
  \begin{center}
    \includegraphics[width=7in]{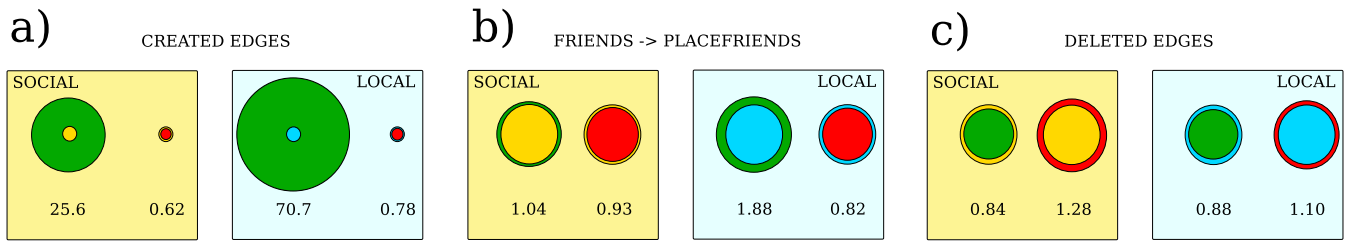}
  \end{center}
  \caption{(colour online) A visualisation of the ratio between the
    actual number of edges formed or deleted (within and between communities represented by green and red circles respectively) and the corresponding
    expected number in the null model (yellow and cyan
   for social and placefriends graphs respectively). The number underneath each pair of nested circles is the value of the represented ratio $r$: the ratio of the actual number of edges within and between communities to the number expected according to the null model. Panel a): the formation
    of social ties within local communities is much stronger than
    expected in the null model (70.7 times larger).
    Panel b): in the
    placefriends graph there is a high probability that a pair of friends in the same community who do not share a place later visit the same place. Panel c): once
    formed, connections are very stable and unlikely to be
    severed with time, to a greater extent within communities than between communities and in local communities in particular.}
    \label{fig:pretty_circles}
\end{figure*}

We first examine which social ties form over the course of the eight
snapshots.  Figure~\ref{fig:pretty_circles}a shows the numbers of
pairs of users who are \textit{not} friends in the first snapshot and
have declared a social tie by the final snapshot. The number of social
ties formed between members of the same social community is $25.6$
times greater than expected when ties form randomly between
disconnected users. The effect is even stronger for local communities:
ties are $70.7$ times more likely than expected to form between
members of the same local community. This shows that social
communities could be useful for applications such as friend
recommendation in services like Gowalla, but local communities might
be even more valuable to consider.

Next, we investigate the conceptually dual problem: which of the
pairs of friends who are \textit{not} placefriends at the start of the
snapshots later become placefriends. Figure~\ref{fig:pretty_circles}b
shows the number of intra- and inter-community pairs of friends who
were not placefriends in the first snapshot, who had become
placefriends by the final snapshot, that is, they were friends in the first snapshot but had not visited the same place, but had visited the same place by the final snapshot.  
Members of the same social
community are only $1.04$ times more likely than expected to become
placefriends than at random. For local communities, the difference is
more pronounced: members of the same local community are $1.88$ times
more likely to become placefriends than at random. One
explanation for this difference could be that if a community is
already focused around physical places, members are more likely to go
to places where other members have been than when the community is
only social.  This again demonstrates the potential benefit of
specifically considering local communities for applications such as
place recommendation.

Finally, we quantify the deletion of social ties between the first
and last snapshots. Figure~\ref{fig:pretty_circles}c shows the number
of social ties that exist in the first snapshot that have been deleted
by the final snapshot. Edge deletion is a comparatively rare event in
online social networks in general, and in Gowalla in particular, with
under $1\%$ of the total edges in the social graph being deleted at
all.  In both social and local communities, edges between members of
the same community are less likely to be deleted than expected at
random, and edges between members of different communities are more
likely than expected to be deleted. Edge deletion in OSNs has not been
extensively studied, due to lack of availability of
data~\cite{Kwak11:Fragile}, but this does seem to suggest that being
in the same community might indicate that a tie in Gowalla is stronger
than one between users in different communities, leading to its
decreased likelihood of deletion.

\subsection{Community events}
The availability of longitudinal data makes it possible to study the
stability of social and local communities, i.e.~to quantify whether
the community decomposition of the graph observed at the beginning
remains stable over time or evolves towards a different partition at
the end.
Other research has generally agreed on the main types of event that
may occur as communities change over
time~\cite{Asur09:Event-based,Greene10:Tracking,Palla07:Quantifying,Tantipathananandh07:Framework}. We
take the definitions of Asur et al.~\cite{Asur09:Event-based} and
denoting the set of nodes making up community $C_k$ in snapshot $i$ by
$C_i^k$, we define the following possible situations:
\begin{itemize}
\item\textbf{{Continue}}: $C_{i+1}^{j}$ is a continuation of $C_i^{k}$ if and
only if $C^{j}_{i+1} = C^{k}_{i}$, i.e. the set of nodes is the same.
\item\textbf{{$\kappa$-Merge}}: $C_{k}^{i}$ and $C_{l}^{i}$ form a merged
community $C_{m}^{i+1}$ if $C_{m}^{i+1}$ contains at least $\kappa\%$
of the nodes belonging to $C_{k}^{i} \cup C_{l}^{i}$, and if it
contains more than half the nodes in each of $C_k^i$ and $C_l^i$.
\item\textbf{{$\kappa$-Split}}: $C_i^{k}$ has been split in snapshot $i+1$ if
$\kappa\%$ of nodes in $C_i^{k}$ are present in different communities
in snapshot $i+1$. For Split and Merge we take $\kappa = 50$ as
in~\cite{Asur09:Event-based}.
\item\textbf{{Form}}: A new community $C_{i+1}^{k}$ forms in snapshot $i+1$ if
none of the nodes in $C_{i+1}^{k}$ were grouped in a community in
snapshot $i$.
\item\textbf{{Dissolve}}: A community $C_i^{k}$ in snapshot $i$ has dissolved in
snapshot $i+1$ if none of the nodes in $C_i^{k}$ are grouped together
in snapshot $i+1$.
\end{itemize}

In order to assess not just whether a community is \textit{exactly the
  same} as in a \textit{Continue} event, but whether it still exists
in some form in the next snapshot although users may have joined or
left, we defined an event \textit{Persist}:\\
\begin{itemize}
\item\textbf{Persist}: $C_{i}^{j}$ persists in snapshot $i+1$ if:
\begin{enumerate}
 \item There is a community $C^{k}_{i+1}$ such that more than half of
 the nodes in $C_{i}^{j}$ are present in $C^{k}_{i+1}$ \item Nodes
 from $C_{i}^{j}$ make up more than half of the nodes in
 $C^{k}_{i+1}$ \end{enumerate}
\end{itemize}
These latter conditions ensure that the majority of the nodes in the
community are still the same, and that the community has not become
merged into a larger one. Note that \textit{Continue} events are a
special case of \textit{Persist} events, in which \textit{all} the
nodes belonging to a given community in a snapshot are put in the same
community in the following snapshot.

\begin{figure*}[tbh]
\centering \includegraphics[width=6.5in]{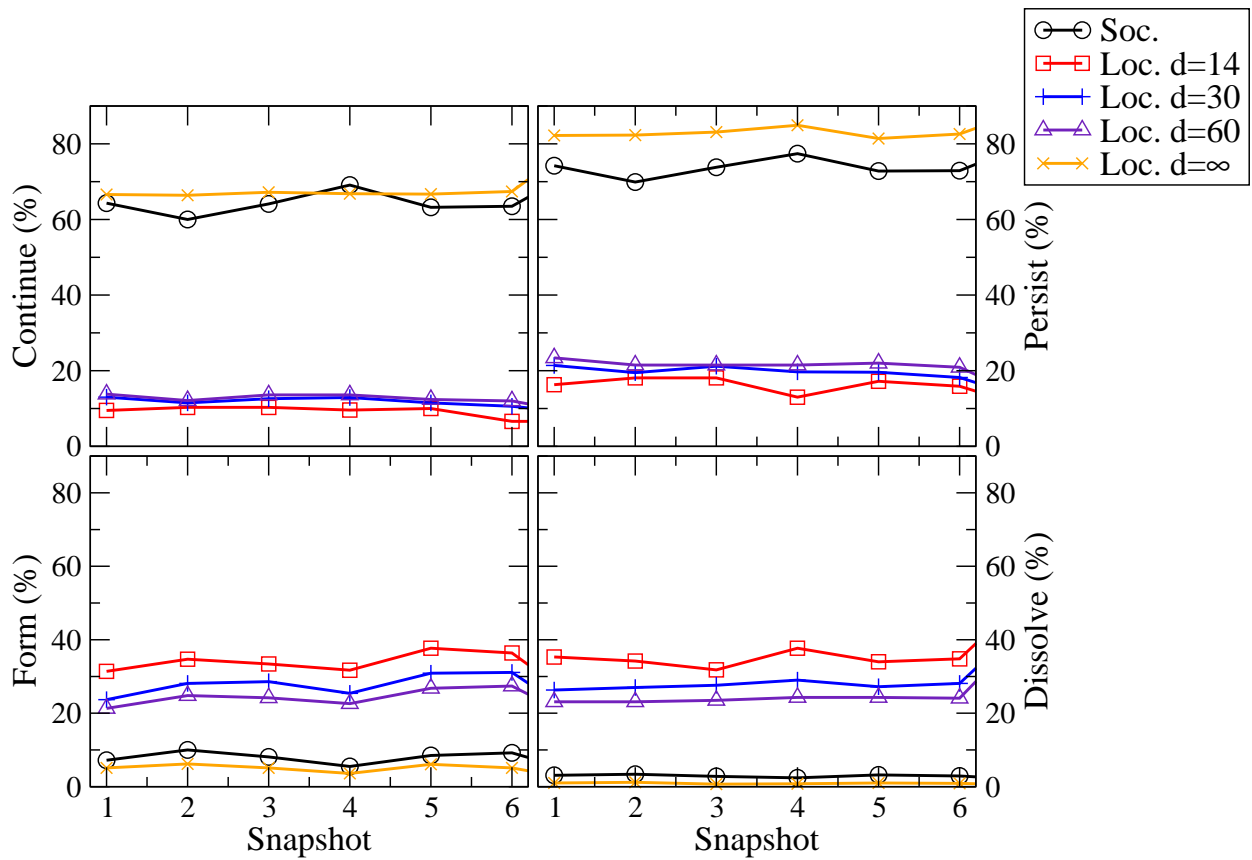}
\caption{(colour online) Percentages of communities undergoing each
  type of event at each snapshot, with placefriends relationships
  lasting $d$ days for the local communities. \textit{Continue} and
  \textit{Persist} events are the more probable ones both in the
  social graph (black line with circles) and in the placefriends graph
  where placefriends ties are indefinitely persistent (orange line with
  crosses). Conversely, when placefriends relationships are restricted
  to common check-ins within 2 weeks, one month and two months
  (resp.~red, blue and purple lines), \textit{Form} and
  \textit{Dissolve} events are more common. This confirms that local
  communities are more volatile and dynamic than social communities.
  Split and Merge events were extremely rare in all cases, and have
  been omitted from the figure.}
\label{fig:events}
\end{figure*}

\subsection{Dynamics of social and local communities}
We have been considering \textit{placefriends} relationships to
continue indefinitely in time: local communities have been obtained
taking users to be placefriends when they have ever checked in to the
same places, regardless of how long ago that was.  We now consider
placefriends relationships to have different lifetimes and examine how
this affects local communities.  Specifically, we study the cases
where users are considered to be placefriends only if they have
checked into one of the same places in a period of $2$ weeks ($14$
days), $1$ month ($30$ days), or $2$ months ($60$ days). Recall that
even though we only have the structure of the social graph during the
snapshot periods, we have check-ins extending back to the beginning of
the service, so it is possible to consider a period of check-ins
before the measurement period began.

We analysed the occurrence of the community events for both social and
local communities over the course of the snapshots.
Figure~\ref{fig:events} shows the percentages of communities in a
snapshot undergoing different types of event, for different time
thresholds for the expiration of a placefriends tie. Note that due to
the definitions of the events, not all communities will undergo one of
the defined events at each snapshot; for example, a community may
break up at the next snapshot, but this will not count as a Dissolve
event if any of the nodes are still in the same community at the next
snapshot.  However, it will not count as a Split event unless enough
nodes of the previous community are present together in a community at
the next snapshot and form more than half of this community; for
example, the case where a $6$-member community breaks into $3$ pairs,
each of which then join different communities, will count neither as a
Dissolve nor a Split, though the community has not Persisted.

Figure~\ref{fig:events} shows that social communities are rather
stable, with a high proportion, $60\%$ or more, remaining unchanged
between pairs of snapshots (Persist) and more than $69\%$ continuing
to exist between each pair of snapshots with minor changes in
membership (Continue). Dissolve events never affect more than $4.5\%$
of communities between a pair of snapshots. Split and Merge events are
very rare, affecting no more than $0.2\%$ of the communities in any
snapshot, and are therefore not shown in the figure.

The figure shows that when we consider placefriends relationships to
have a limited duration, local communities become highly volatile:
when users must have checked in to the same place within the past $2$
weeks to be considered placefriends, under $20\%$ of local communities
persist from snapshot to snapshot. The proportion remains under $25\%$
even when the duration is increased to $2$ months, which is quite a
generous period of time. This is in stark contrast to the very high
proportions of persistent communities observed when placefriends
relationships are assumed to continue indefinitely; in that case,
local communities are \textit{more} stable than social communities,
with over $80\%$ of communities in any one snapshot persisting to the
next, and below $2\%$ dissolving.

The instability of local communities when placefriends edges have
limited duration may be due to users not consistently checking in at
locations when they go there, rather than their ceasing to go to the
same places. This would reflect how people use Gowalla as a service,
rather than their true mobility, or indeed their relationships with
their online friends. Just because users have not checked in at the
same place for a while, it does not necessarily mean they no longer
see one another, that they are no longer friends, or that they no
longer visit that place, and so we must be careful what we infer from
the instability of these communities. To investigate what is happening
in more detail, we examined firstly whether or not users in local
communities that dissolved between snapshots had stopped using the
service, and secondly, if they had not, whether they were still
checking in to places in the same area. We found that in most cases,
users continued to make check-ins in the same geographic area as they
had been previously.  This would indicate that the dissolution of the
local community probably does not indicate that they are no longer in
the same area.  Furthermore, for all of the $14$-, $30$- and $60$-day
lifespans of placefriends relationships, between $30\%$ and $35\%$ of
the communities that dissolved after the first snapshot reappeared in
one of the later snapshots. This may indicate that the users are still
visiting the places that they have in common, but are not regularly
checking in using Gowalla. Previous research by Lindqvist et
al.~\cite{Lindqvist11:Mayor} into how people use Foursquare, another
location-based social network similar to Gowalla, found that people
had many reasons for not checking in at locations, ranging from
privacy concerns to the fact that they found it `boring' to keep on
checking in repeatedly. The unstable local communities that we see
here may well be a consequence of this type of behaviour.

\section{Conclusions}
We have analysed social and local communities in Gowalla, an online
location-based social network, and demonstrated that the two community
structures do not yield the same user groupings.  Despite the tendency
of spatially close users to form social connections, systems that aim
to make use of the existence of both types of community may not be
able to rely on simply considering the community structure of the
social network, but should take geography into account explicitly. We
have seen that local communities could be valuable for friend
suggestion and place recommendation, since edges are more likely to
form within local communities than within social communities, and
friends in the same local community are more likely to visit the same
places. Finally, we have shown that while the social graph changes
slowly and thus social communities are quite stable, local communities
can be very transient or very stable depending on the lifetime given
to the placefriends relationship. This has implications for systems
aiming to make use of local communities: the choice of timescale at
which the placefriends relationship is considered may be crucial due
to the way in which users perform check-ins.

These results suggest that location-aware applications aiming to
exploit the existence of community structure in OSNs should not rely
only on the detection of social communities: these communities can
fail to capture local groups.  By taking geographic information into
account, local communities can be extracted and these may be more
useful than social communities in applications such as providing
personalised friend suggestions or place recommendations. However,
systems making use of local communities should carefully choose the
timescale at which they perform community detection, according to the
particular needs of the application. This work makes a step towards
online social services and systems being able to make better use of
community information, as they become increasingly location-aware.

\small{ C. B. is a recipient of the Google Europe Fellowship
in Mobile Computing, and this research is supported in part by this
Google Fellowship. V.N. acknowledges support from the EPSRC project
MOLTEN (EP/l017321/1) and from the EU project LASAGNE, contract
No. 318132 (STREP).}


\begin{thebibliography}{2}%
\makeatletter
\providecommand \@ifxundefined [1]{%
 \@ifx{#1\undefined}
}%
\providecommand \@ifnum [1]{%
 \ifnum #1\expandafter \@firstoftwo
 \else \expandafter \@secondoftwo
 \fi
}%
\providecommand \@ifx [1]{%
 \ifx #1\expandafter \@firstoftwo
 \else \expandafter \@secondoftwo
 \fi
}%
\providecommand \natexlab [1]{#1}%
\providecommand \enquote  [1]{``#1''}%
\providecommand \bibnamefont  [1]{#1}%
\providecommand \bibfnamefont [1]{#1}%
\providecommand \citenamefont [1]{#1}%
\providecommand \href@noop [0]{\@secondoftwo}%
\providecommand \href [0]{\begingroup \@sanitize@url \@href}%
\providecommand \@href[1]{\@@startlink{#1}\@@href}%
\providecommand \@@href[1]{\endgroup#1\@@endlink}%
\providecommand \@sanitize@url [0]{\catcode `\\12\catcode `\$12\catcode
  `\&12\catcode `\#12\catcode `\^12\catcode `\_12\catcode `\%12\relax}%
\providecommand \@@startlink[1]{}%
\providecommand \@@endlink[0]{}%
\providecommand \url  [0]{\begingroup\@sanitize@url \@url }%
\providecommand \@url [1]{\endgroup\@href {#1}{\urlprefix }}%
\providecommand \urlprefix  [0]{URL }%
\providecommand \Eprint [0]{\href }%
\providecommand \doibase [0]{http://dx.doi.org/}%
\providecommand \selectlanguage [0]{\@gobble}%
\providecommand \bibinfo  [0]{\@secondoftwo}%
\providecommand \bibfield  [0]{\@secondoftwo}%
\providecommand \translation [1]{[#1]}%
\providecommand \BibitemOpen [0]{}%
\providecommand \bibitemStop [0]{}%
\providecommand \bibitemNoStop [0]{.\EOS\space}%
\providecommand \EOS [0]{\spacefactor3000\relax}%
\providecommand \BibitemShut  [1]{\csname bibitem#1\endcsname}%
\let\auto@bib@innerbib\@empty
\bibitem [{Note1()}]{Note1}%
  \BibitemOpen
  \bibinfo {note} {The exponents of the tails of the degree distributions have
  been computed using the maximum likelihood estimator (MLE) for discrete
  power-law distributions, as indicated in Ref.~\cite {Clauset2007a}. The
  values of $k_{min}$ corresponding to the maximum likelihood fit are
  $k_{min}=79$ and $k_{min}=67$, respectively for social and placefriend
  graphs.}\BibitemShut {Stop}%
\bibitem [{Note2()}]{Note2}%
  \BibitemOpen
  \bibinfo {note} {The exponents of the tails of the distributions of community
  sizes have been computed using the maximum likelihood estimator (MLE) for
  discrete power-law distributions, as indicated in Ref.~\cite {Clauset2007a}.
  The fit was perfomed by setting $s_{min}=30$.}\BibitemShut {Stop}%
\end{thebibliography}%


\begin{thebibliography}{10}
\bibitem{Barabasi2002rev} R. Albert and A.-L. Barabasi,
  \textit{Rev. Mod. Phys.} \textbf{74}, 47 (2002).

\bibitem{Newman2003rev} M. E. J. Newman, \textit{SIAM Review}
  \textbf{45}, 167-256 (2003).
  
 \bibitem{Boccaletti2006} S. Boccaletti, V. Latora, Y. Moreno,
   M. Chavez and D.-U. Hwang, \textit{Phys. Rep.} \textbf{424}, 175 -
   308 (2006).

\bibitem{Barrat2008} A. Barrat, M. Barth\'{e}lemy and A. Vespignani,
  \textit{Dynamical processes on complex networks} (2008).
  
 \bibitem{Newman2010} M. Newman, \textit{Networks: an introduction} (2010).
 
\bibitem{granovetter_1973} Granovetter, M., \textit{Am. J. Sociol.}
  \textbf{78}, 1360-1380 (1973).

\bibitem{coleman_1988} Coleman, J. S., \textit{Am. J. Sociol.}
  \textbf{94}, S95-S120 (1988).

\bibitem{Castellano2009} C. Castellano, S. Fortunato and V. Loreto,
  \textit{Rev. Mod. Phys.} \textbf{81,}, 591-646 (2009).

\bibitem{Szell2010} M. Szell, R. Lambiotte and S. Thurner,
  \textit{Proc. Natl. Acad. Sci. USA} \textbf{107}, 31,
  13636--13641(2010).
  
\bibitem{Onnela2010} J.-P. Onnela and F. Reed-Tsochas,
  \textit{Proc. Natl. Acad. Sci. USA} \textbf{107}, 18375-18380
  (2010).
  
\bibitem{Lewis2012} K. Lewis, M. Gonz\'alez and J. Kaufman,
  \textit{Proc. Natl. Acad. Sci. USA} \textbf{109}, 68-72 (2012).
  
\bibitem{Borge-Holthoefer2011} J. Borge-Holthoefer, A. Rivero,
  Garc{\'\i}a, I, Cauh{\'e}, Elisa, A. Ferrer, D. Ferrer, D. Francos, D. I{\~n}iguez,
  M. P. P{\'e}rez, G. Ruiz, F. Sanz, F. Serrano, C. Vi{\~n}as, A. Taranc{\'o}n and
  Y. Moreno, \textit{PLoS ONE} \textbf{6} (2011).
  
 \bibitem{Crandall10:Inferring} { Crandall, D.~J., Backstrom, L.,
   Cosley, D., Suri, S., Huttenlocher, D., and Kleinberg, J.}
   \newblock {\em Proc. Natl. Acad. Sci. USA} \textbf{107}, 52 (2010),
   22436--22441.
  
\bibitem{Jones2013} J. J. Jones, J. E. Settle, R. M. Bond,
  C. J. Fariss, C. Marlow and J. H. Fowler, \textit{PLoS ONE}
  \textbf{8} (2013).
  
\bibitem{Liu12:Simplifying}
{ Liu, Y., Viswanath, B., Mondal, M., and Mislove, A.}
\newblock {\em Proceedings of WWW '12} (2012).

\bibitem{Papadopoulos11:Community}
{ Papadopoulos, S., Kompatsiaris, Y., Vakali, A., and Spyridonos, P.},
\newblock {\em Data Mining and Knowledge Discovery}, 1--40 (2011).

\bibitem{Pujol10:Little}
{ Pujol, J.~M., Erramilli, V., Siganos, G., Yang, X., Laoutaris, N.,
  Chhabra, P., and Rodriguez, P.},
\newblock {\em Proceedings of SIGCOMM '10}, 375--386 (2010).

\bibitem{Girvan2002} M. Girvan and M. E. J. Newman,
  \textit{Proc. Natl. Acad. Sci. USA} \textbf{99}, 7821-7826 (2002).

\bibitem{Newman2004a} M. E. J. Newman, \textit{Eur. Phys. J. B}
  \textbf{38}, 321--330 (2004).
  
\bibitem{Newman2006} M. E. J. Newman, \textit{Proc. Natl. Acad. Sci.
  USA} \textbf{103}, 8577-8582 (2006).
  
\bibitem{Fortunato2009} S. Fortunato, \textit{Phys. Rep.}
  \textbf{486,} 75-174 (2009).

\bibitem{Backstrom06:Group}
{ Backstrom, L., Huttenlocher, D., Kleinberg, J., and Lan, X.},
\newblock{\em Proceedings of KDD '06}, 44--54 (2006).
  
\bibitem{Kumar10:Structure}
{ Kumar, R., Novak, J., and Tomkins, A.},
\newblock {\em Link Mining: Models, Algorithms, and Applications}, P.~S.~S.
  Yu, J.~Han, and C.~Faloutsos, Eds. Springer New York, 337--357 (2010).
  

\bibitem{Mislove07:Measurement}
{ Mislove, A., Marcon, M., Gummadi, K.~P., Druschel, P., and Bhattacharjee,
  B.},
\newblock {\em Proceedings of IMC '07}, 29--42 (2007).

\bibitem{Facebook12:Building} { Facebook}, \newblock
  \url{http://developers.facebook.com}\\\url{/blog/post/2012/03/07/building-be%
    tter-stories-}\\\url{with-location-and-friends}, (2012).

\bibitem{Twitter09:Think}
{ Twitter},
\newblock
\url{http://blog.twitter.com/2009/11/}\\\url{think-globally-tweet-locally.html}
(2009).
  
\bibitem{Foursquare13:What} {Foursquare}, \newblock
  \url{http://blog.foursquare.com/2013/01/17/}\\\url{what-the-last-500000000-check-ins-look}\\ \url{-like-and-what-they-show-about}\\\url{-the-future-of-foursquare/},
  (2013).

\bibitem{Backstrom10:Find}
{ Backstrom, L., Sun, E., and Marlow, C.},
\newblock {\em Proceedings of WWW '10}, 61--70 (2010).

\bibitem{Cranshaw10:Bridging}
{ Cranshaw, J., Toch, E., Hong, J., Kittur, A., and Sadeh, N.},
\newblock{\em Proceedings of UbiComp '10}, 119--128 (2010).

\bibitem{Liben-Nowell05:Geographic}
{ Liben-Nowell, D., Novak, J., Kumar, R., Raghavan, P., and Tomkins, A.},
\newblock {\em Proc. Natl. Acad. Sci. USA} \textbf{102}, 33,
11623--11628 (2005).
  
\bibitem{Scellato11:Socio-spatial}
{ Scellato, S., Noulas, A., Lambiotte, R., and Mascolo, C.},
\newblock {\em Proceedings of ICWSM '11} (2011).

\bibitem{Clauset2007a} A. Clauset, C. R. Shalizi and M. E. J. Newman,
  \textit{SIAM Review} \textbf{51}, 661-703 (2007).

\bibitem{Newman2002} M. E. J. Newman, \textit{Phys. Rev. Lett.}
  \textbf{89,}, 208701 (2002).

\bibitem{Newman2003} M. E. J. Newman, \textit{Phys. Rev. E}
  \textbf{67,}, 026126 (2003).

\bibitem{Blondel08:Fast} { Blondel, V.~D., Guillaume, J.-L.,
  Lambiotte, R., and Lefebvre, E.}, \newblock {\em J. Stat. Mech.}
  \textbf{10}, P10008 (2008).

\bibitem{Arenas2007} A. Arenas, J. Duch, A. Fernández and S. Gómez,
  \textit{New J. Phys.} \textbf{9}, 176 (2007).

\bibitem{Brown12:Importance} { Brown, C., Nicosia, V., Scellato, S.,
  Noulas, A., and Mascolo, C.}, \newblock {\em Proceedings of WOSN
  '12\/} (2012).

\bibitem{Kwak09:Mining} { Kwak, H., Choi, Y., Eom, Y.-H., Jeong, H.,
  and Moon, S.}, \newblock {\em Proceedings of IMC '09\/}, 301--314
  (2009).


\bibitem{Onnela11:Geographic}
{ Onnela, J.-P., Arbesman, S., Gonz{\'a}lez, M.~C., Barab{\'a}si, A.-L., and
  Christakis, N.~A.},
\newblock {\em PLoS ONE \textbf{6}}, 4 (2011).


\bibitem{Scellato11:Exploiting}
{ Scellato, S., Noulas, A., and Mascolo, C.},
\newblock {\em Proceedings of KDD '11\/}, 1046--1054 (2011).

\bibitem{Stewart41:Inverse}
{ Stewart, J.~Q.},
\newblock {\em Science \textbf{93}}, 2404, 89--90 (1941).

\bibitem{Kwak11:Fragile}
{ Kwak, H., Chun, H., and Moon, S.},
\newblock {\em Proceedings of CHI '11\/}, 1091--1100 (2011).

\bibitem{Asur09:Event-based}
{ Asur, S., Parthasarathy, S., and Ucar, D.},
\newblock {\em ACM Trans. Knowl. Discov. Data} \textbf{3}, 4 (2009).

\bibitem{Greene10:Tracking}
{ Greene, D., Doyle, D., and Cunningham, P.},
\newblock {\em International Conference on Advances in Social Networks
  Analysis and Mining (ASONAM)\/}, 176 --183, (2010).

\bibitem{Palla07:Quantifying} { Palla, G., Barab\'asi, A.-L., and
  Vicsek, T.}, \newblock {\em Nature} \textbf{446}, 7136, 664--667
  (2007).

\bibitem{Tantipathananandh07:Framework}
{ Tantipathananandh, C., Berger-Wolf, T., and Kempe, D.},
\newblock {\em Proceedings of KDD '07}, 717--726 (2007).

\bibitem{Lindqvist11:Mayor}
{ Lindqvist, J., Cranshaw, J., Wiese, J., Hong, J., and Zimmerman, J.},
\newblock {\em Proceedings of CHI '11\/}, 2409--2418 (2011).

\end{thebibliography}
\end{document}